\documentclass[12pt]{article}
\usepackage{a4,amsmath}
\usepackage{epsfig}
\begin{document}
\begin{center}
\begin{Large}
{\bf{Dynamic rearrangement of vacuum\\ and the phase transitions\\[2mm] 
 in the geometric structure of space-time}} 
\end{Large}   
\vskip 5mm
\begin{large}
      George Yu. Bogoslovsky \\
\end{large}
\vskip 3mm
{\footnotesize{{\it{Skobeltsyn Institute of Nuclear Physics, Moscow State University, 119992 Moscow, Russia\\
E-mail\,:}} bogoslov@theory.sinp.msu.ru}}\\
\end{center}
\vskip 9mm
\begin{small}
\noindent
{\bf{Abstract.}} \,\,It is shown that in the case of spontaneous breaking of the original gauge symmetry, a dynamic rearrangement of vacuum may lead to the formation of some  anisotropic condensates. The appearance of such condensates causes the respective phase transitions in the geometric structure of space-time and creates a flat anisotropic, i.e. Finslerian event space.
Actually there arises either a flat relativistically-invariant Finslerian space with partially broken 3D isotropy, i.e. axially-symmetric  space, or a flat relativistically-invariant Finslerian space with entirely broken 3D isotropy. The fact that any entirely anisotropic relativistically-invariant Finslerian event space belongs to a 3-parameter family of such spaces gives rise to a fine structure
of the respective geometric phase transitions. In the present work the fine structure of the geometric phase transitions is studied by classifying all the metric states of the entirely anisotropic event space and the respective  mass shell equations.\\[3mm]
\end{small} 
{\footnotesize{   
\noindent
{\it{Keywords}}\,: Lorentz and gauge symmetry; \,Spontaneous symmetry breaking; \,Finslerian space-time\\[1mm]
\noindent
{\it{2000 Mathematics Subject Classification}}\,: 53C60; 53C80; 83A05; 81T13; 81R40}}\\[10mm]                 
\noindent
{\large{\bf 1\,. Introduction}}

\bigskip

\noindent
In terms of general relativity, the space-time is known to be Riemannian. According to the Einstein equations, the  distribution and motion of matter determine only the space-time curvature and have never affect the geometry of tangent spaces. In other words, irrespective of the distribution and properties of the material medium that fills the Riemannian space-time, any flat tangent space-time remains the space-time of special relativity, i.e. Minkowski space.

Generally speaking, a flat space-time does not exhibit  Minkowski geometry in all cases. Such geometry arises only if a given flat space-time admits the 6-parameter Lorentz group to be a homogeneous isometry group. The 6-parameter Lorentz group is known to include the 3-parameter Lorentz boosts and the 3D rotation subgroup. If, however, the isotropy of 3D space is broken in a way, the flat space-time metric can no longer be described by the quadratic form of the coordinate differentials, but is a certain (and, generally speaking, arbitrary enough) homogeneous function of the differentials of  degree two. In this case, the flat space-time is said to have the Finslerian geometry [1],\,[2].

Attempts have long been made (see [3]--[8] and  the relatively new interesting work [9] for instance) to generalize the field theory
and the Einstein equations for the Finslerian space-time. That task is difficult because, first of all, the Finslerian metric tensor depends on not only the base manifold points, but also the geometric objects of, generally, arbitrary nature. Therefore, any noticeable relevant advance in the field involves additional physical concepts. In particular, the extremely fruitful concept of Lorentz symmetry violation at relative velocities of inertial reference frames in immediate proximity to light velocity should be noted. The concept was suggested in [10],\,[11] and led finally to construction of the viable model for flat space-time with partially broken 3D isotropy [12]. The model is described by the following Finslerian metric\,:
\begin{equation}\label{1}
ds^2=\left[\frac{(dx_0-\boldsymbol\nu d\boldsymbol x)^2}{dx_0^2-d\boldsymbol x^{\,2}}\right]^r
(dx_0^2-d\boldsymbol x^{\,2})\,,
\end{equation}
where the unit vector $\,\boldsymbol\nu\,$ indicates a preferred direction in 3D space; the dimensionless parameter $\,r\,$  determines the anisotropy magnitude, i.e. the degree of deviation of the Finslerian metric (1) from the metric of  isotropic Minkowski space. In this case, the Minkowski metric is clearly the ultimate case of the metric (1) at $\,r=0\,.$

Another ultimate case ($\,r=1\,$) is also of importance when constructing a consistent dynamics pattern for space-time manifold. In this case, according to (1), the metric $\,ds\,$ degenerates into a total differential of absolute time. This transformation of the metric leads us to conclude that some phase transitions can occur in the space-time geometric structure and may be related to the phase transitions arising in the system of  interacting fundamental fields in case the gauge symmetry breaks spontaneously. We shall discuss this aspect in more detail below; but would pay attention now to another important circumstance that concerns the specific form of the Finslerian metric (1).

Any of the Finslerian metrics 
$\,\,ds^2=f((dx_0-\boldsymbol\nu d\boldsymbol x)^2/(dx_0^2-d\boldsymbol x^{\,2}))(dx_0^2-d\boldsymbol x^{\,2})\,$
\, (where $\,f((dx_0-\boldsymbol\nu d\boldsymbol x)^2/(dx_0^2-d\boldsymbol x^{\,2}))\,$ is, in many respects, an arbitrary function of its own argument) also describes a certain flat Finslerian space-time with partially broken 3D isotropy, i.e. an axially symmetric Finslerian space. At the same time, if and only if $\,f\,$ is of the form $f=((dx_0-\boldsymbol\nu d\boldsymbol x)^2/(dx_0^2-d\boldsymbol x^{\,2}))^r,$ the respective metric (1) will describe the flat anisotropic space-time, which permits not only the 1-parameter group of rotations about vector $\,\boldsymbol\nu\,,$ but also the homogeneous 3-parameter group of isometries that consists of only noncompact transformations. Such transformations link the physically equivalent inertial reference frames in the anisotropic space-time (1) and are called the generalized Lorentz transformations, or the generalized Lorentz boosts. As a result, we may assert that, when going over from  Minkowski space to the Finslerian space (1) with partially broken 3D isotropy, Lorentz space-time symmetry proves to be also broken, but the relativistic symmetry represented by the group of generalized Lorentz boosts remains valid [13]--[21]. The respective theory developed in [13]--[21] is called Anisotropic Special Relativity. However, after [22],\,[23], where such a theory has been widely tested, it is increasingly referred to as General Very Special Relativity (GVSR). The same can be said in respect to the 8-parameter inhomogeneous group of isometries of the Finslerian space-time (1): this group is increasingly referred to as DISIM${_{b}} $(2).

In terms of the above described Finslerian model, the anisotropy of the flat space-time is produced by the relativistically-invariant axially-symmetric fermion-antifermion condensate formed under spontaneous breaking of the initial gauge symmetry and when the fundamental matter fields acquire masses. Contrary to the standard Higgs mechanism and to its alternative pattern [24],\,[25], which treats the scalar fermion-antifermion condensate instead of the Higgs condensate, the vacuum rearrangement accompanied by formation of the relativistically-invariant axially-symmetric fermion-antifermion condensate leads to changing the flat space-time geometry, namely, the Finslerian geometry with the metric (1) replaces Minkowski geometry. In this case, as noted above, such a geometric phase transition preserves the relativistic symmetry, but violates the Lorentz symmetry of the theory.

Lately, another (string-motivated) approach to the problem of breaking the Lorentz symmetry is developed along with the Finslerian approach. The fact is that, even if a base unified theory exhibits  Lorentz symmetry at the most fundamental level, that symmetry can be broken spontaneously due to formation of the condensate of the vector or (for instance) tensor field. The assumed occurrence of such a condensate, or a constant classical field against Minkowski space background implies that it may affect the dynamics of the fundamental fields and, thereby, modify the Standard Model of strong, weak, and electromagnetic interactions. Since the constant classical field is transformed under passive transformations as a Lorentz vector or tensor, this effect will properly be allowed for by extension of the Standard Model Lagrangian using the additional terms, which are every possible Lorentz-covariant convolutions of the condensate with the standard fundamental fields. The phenomenological theory based on the given Lorentz-covariant modification of the Standard Model was called the Standard Model Extension (SME) [26]--[32]. By its construction, that model is not Lorentz-invariant because its Lagrangian fails to remain invariant under active Lorentz transformations of the fundamental fields against the background of fixed condensate. Additionally, in the SME context, Lorentz symmetry violation with respect to the active Lorentz transformations implies also the relativistic symmetry violation because the presence of non-invariant condensate breaks the physical equivalence of the various inertial reference frames.

Of course, it cannot be excluded that the Nature is so organized that, on the Planck energy scales, not only Lorentzian symmetry, but also the above-described generalized Lorentzian symmetry, will prove to be broken either entirely or partially [33]. Even in this case, however, Finslerian geometric space-time model may prove to be more adequate compared with the Riemannian model. Although the like viewpoint has been known (see [34]--[44] for instance), it is apt to note that the absence of some local isometry group in the Finslerian space-time necessitates additional physical criteria that make it possible to select only those Finslerian metrics from their set, which permit description of the geometric properties of space-time manifold. For example, the authors of [45]--[47] used availability of the conformal and projective structures of a Finslerian space as such criteria to show that the Finslerian space-time that satisfies the criteria must be Berwald's special Finslerian space.

Returning to the Finslerian spaces that permit the homogeneous noncompact 3-parameter isometry groups and, hence, have the relativistic symmetry, the present work will pay main attention below to further investigating the flat Finslerian event spaces with entirely broken 3D isotropy.

\vspace*{10mm}

\noindent
{\large{\bf 2\,. Classification of all metric states of the flat entirely anisotropic \phantom{eve}event space}}
\vspace*{5mm}

\noindent
In terms of the relativity theory, the basic property of axially-symmetric Finslerian space-time (1) is that the latter is also relativistically symmetric. In other words, the transformations which link the various inertial reference frames belong to the group of its isometries and, in their turn, form a separate 3-parameter group.
As to the axial symmetry, it means that under transition from Minkowski space to the Finslerian space-time (1), the 3D space isotropy is broken but partially.

Apart from the flat axially-symmetric Finslerian space-time (1), which is created by axially-symmetric relativistically-invariant fermion-antifermion condensate, the flat entirely anisotropic Finslerian space-time, which is created by entirely  anisotropic relativistically-invariant three-gluon condensate\,\footnote{A feasibility of forming the three-gluon condensate
ensues from work [48] that demonstrates a
spontaneous generation of three-gluon gauge invariant ef\/fective
interaction.}, may arise in the course of the vacuum rearrangement.  The most general form of the respective entirely anisotropic Finslerian metric has been found in [49] and proved to depend on three dimensionless parameters $\,r_1\,,\ r_2\,$ and $\,r_3\,$ and to be presented as
\begin{equation}\label{2}
\begin{array}{rl}
ds=&(dx_0-dx_1-dx_2-dx_3)^{(1+r_1+r_2+r_3)\,/\,4}
(dx_0-dx_1+dx_2+dx_3)^{(1+r_1-r_2-r_3)\,/\,4}\\
\times &(dx_0+dx_1-dx_2+dx_3)^{(1-r_1+r_2-r_3)\,/\,4}
(dx_0+dx_1+dx_2-dx_3)^{(1-r_1-r_2+r_3)\,/\,4}\,.
\end{array}
\end{equation}
The range of admissible values of $\,r_1\,,\ r_2\,$ and $\,r_3\,$ is restricted by the conditions
$$
\begin{array}{ll}
1+r_1+r_2+r_3\ge 0\,,&1+r_1-r_2-r_3\ge 0\,,\\
1-r_1+r_2-r_3\ge 0\,,&1-r_1-r_2+r_3\ge 0
\end{array}
$$
and has the form of the regular tetrahedron $\,A\,B\,C\,D,\,$ shown in Fig.\,1.\\ 
\begin{figure}[hbt]
\begin{center}
\epsfig{figure=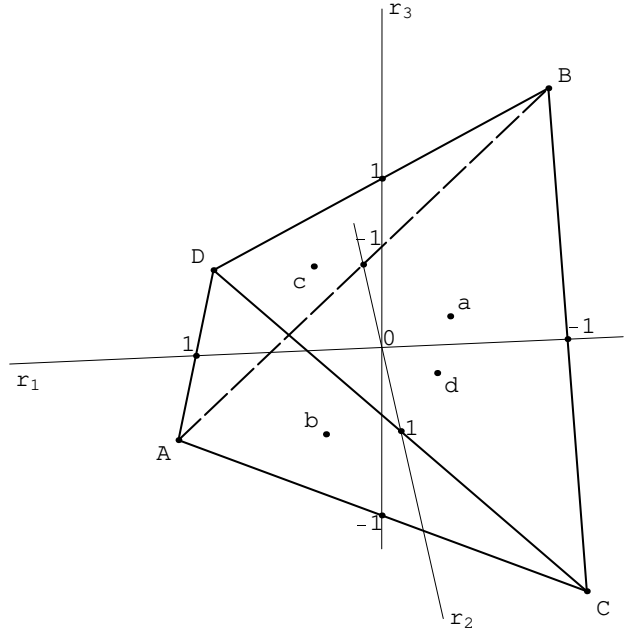,width=10.29cm,height=9.5cm}
\end{center}
\caption{The range of admissible values of $\,r_1\,,\ r_2\,$ and $\,r_3\,.$}
\end{figure}

\noindent
Thus the metric (2) describes 3-parameter family of 
relativistically-invariant Finslerian spaces with entirely broken 3D isotropy.
At $\,r_1=r_2=r_3=0\,$, the metric (2) reduces to the fourth root of the product of four 1-forms\,:
\begin{equation}\label{3}
\begin{array}{rl}
ds_{_{B-M}}=&[\,(dx_0-dx_1-dx_2-dx_3)(dx_0-dx_1+dx_2+dx_3)\\
\times &\phantom{[\,}(dx_0+dx_1-dx_2+dx_3)(dx_0+dx_1+dx_2-dx_3)\,]^{\,1/4}\,.\\
\end{array}
\end{equation}
If the new coordinates $\,{\xi}_i\,$	 are introduced so that                                                                                      
$$
{\xi}_i=A_{ij}x_j\,,\qquad A_{ij}=\left(
\begin{array}{rrrr}
1&-1&-1&-1\\
1&-1&1&1\\
1&1&-1&1\\
1&1&1&-1
\end{array}
\right)\,,
$$
the expression (3) in these coordinates will take the standard form of the Berwald-Mo\'ore metric [50],\,[51], i.e. $ds_{_{B-M}}=\root 4\of{{\xi}_1{\xi}_2{\xi}_3{\xi}_4}\,.$ So, we see that the Berwald-Mo\'or metric presented by expression (3) is the particular case of metric (2) at $\,r_1=r_2=r_3=0\,$ (this is the central point of tetrahedron $\,A\,B\,C\,D\,$ in Fig.\,1)\,.

The tetrahedron vertex $\,A\,$ is in correspondence with the $\,r_{\alpha}\,$ values $\,(r_1=1\,, r_2=-1\,, r_3=-1)\,,$  vertex $\,B\,$ with $\,(r_1=-1\,, r_2=-1\,, r_3=1)\,,$ vertex $\,C\,$ with $\,(r_1=-1\,, r_2=1\,, r_3=-1)\,,$ and vertex $\,D\,$ with $\,(r_1=1\,, r_2=1\,, r_3=1)\,.$ At each of the vertices, the metric (2) that describes the space-time with entirely broken 3D isotropy degenerates into the respective 1-form, i.e. into the total differential of absolute time\,:
$$
\begin{array}{ll}
ds_{_A}=dx_0-dx_1+dx_2+dx_3\,;&\ ds_{_B}=dx_0+dx_1+dx_2-dx_3\,;\\
ds_{_C}=dx_0+dx_1-dx_2+dx_3\,;&\ ds_{_D}=dx_0-dx_1-dx_2-dx_3\,.\\
\end{array}
$$
If this observation is now compared with the above mentioned fact that the metric (1) describing space-time with partially broken 3D isotropy gets also degenerated at $\,r=1\,$ into the total differential of absolute time, the idea arises [49] that the absolute time is not a stable degenerate state of space-time and may turn into either partially anisotropic space-time (1) or entirely anisotropic space-time (2). In any case, the respective  geometric phase transition from the absolute time to 4D space-time may be treated to be an Act of Creation of 3D space. This phenomenon is accompanied by rearrangement of the vacuum state of the system of interacting fundamental fields, resulting in that elementary particles acquire masses. In the case of space-time (1)
the acquired mass of a particle is specified by the tensor 
\begin{equation}\nonumber 
m_{\alpha\beta}=m(1-r)({\delta}_{\alpha\beta}+r{\nu}_\alpha\,{\nu}_
\beta )\,.              
\end{equation}
As for the space-time (2), the acquired mass is specified by the tensor
\begin{equation}\nonumber
m_{\alpha\beta} = 
m\left( \begin{array}{ccc}
(1-{r_1}^2) & (r_3-r_1r_2) & (r_2-r_1r_3) \\
(r_3-r_1r_2) & (1-{r_2}^2) & (r_1-r_2r_3) \\
(r_2-r_1r_3) & (r_1-r_2r_3) & (1-{r_3}^2)
\end{array} \right)\,.
\end{equation} 
Only after the above described process is complete, the concepts of spatial extension and of reference frame get physically meaningful; (in a massless world, a spatial extension of anything, as well as one or another reference frame, is meaningless to speak of\,\footnote{It should be noted that as early as in one of the first unified gauge theories (namely, the conformal Weyl theory [52],\,[53]) the very notion of space-time interval gets physically meaningful only after violation of local conformal symmetry and after the initially massless Abelian vector gauge field acquires mass.}). Finally, attention should be paid also to the fact that, formally, the absolute time serves as the connecting link, via which the correspondence principle is satisfied for the Finslerian spaces with partially and entirely broken 3D isotropy.

With the view to  studying the fine structure of the geometric phase transitions, it is expedient to examine some Finslerian metrics obtainable using the metric (2) as generatrix and selecting the appropriate characteristic subsets from the total set of the admissible values of the parameters $\,r_1\,,\ r_2\,$ and $\,r_3\,.$

According to Fig. 1, the relation $\,r_3=-1-r_1-r_2\,$ holds among the parameters $\,r_{\alpha}\,$ on the $\,A\,B\,C\,$ face. Therefore, we use (2) to obtain
\begin{equation}\label{4}
\begin{array}{rl}
ds_{_{ABC}}=&(dx_0-dx_1+dx_2+dx_3)^{(1+r_1)/2}(dx_0+dx_1-dx_2+dx_3)^{(1+r_2)/2}\\
\times &(dx_0+dx_1+dx_2-dx_3)^{-(r_1+r_2)/2}\,.\\
\end{array}
\end{equation}
At the central point $\,d\,$ of face $\,A\,B\,C\,,$ the parameters $\,r_{\alpha}\,$ are	$\,r_1=r_2=r_3=-1/3\,,$ while (4) reduces to the cubic root
\begin{equation}\label{5}
ds_{_d}=\root 3 \of {(dx_0-dx_1+dx_2+dx_3)(dx_0+dx_1-dx_2+dx_3)(dx_0+dx_1+dx_2-dx_3)}\,.
\end{equation}
On the $\,B\,C\,D\,$ face, the relation $\,r_3=1+r_1-r_2\,$ 	holds among the parameters $\,r_{\alpha}\,.$ Accordingly, formula (2) gives
\begin{equation}\label{6}
\begin{array}{rl}
ds_{_{BCD}}=&(dx_0-dx_1-dx_2-dx_3)^{(1+r_1)/2}(dx_0+dx_1+dx_2-dx_3)^{(1-r_2)/2}\\
\times &(dx_0+dx_1-dx_2+dx_3)^{-(r_1-r_2)/2}\,.\\
\end{array}
\end{equation}
At the central point $\,a\,$ of face $\,B\,C\,D\,, \ r_1=-1/3\,,\,r_2=r_3=1/3\,$ and, according to (6), we again obtain the metric in the form of cubic root\,:
\begin{equation}\label{7}
ds_{_a}=\root 3 \of {(dx_0-dx_1-dx_2-dx_3)(dx_0+dx_1-dx_2+dx_3)(dx_0+dx_1+dx_2-dx_3)}\,.
\end{equation}
On the $\,A\,B\,D\,$ face, $\,r_3=1-r_1+r_2\,,$ resulting in
\begin{equation}\label{8}
\begin{array}{rl}
ds_{_{ABD}}=&(dx_0+dx_1+dx_2-dx_3)^{(1-r_1)/2}(dx_0-dx_1-dx_2-dx_3)^{(1+r_2)/2}\\
\times &(dx_0-dx_1+dx_2+dx_3)^{(r_1-r_2)/2}\,.\\
\end{array}
\end{equation}
At the central point $\,c\,$ of face $\,A\,B\,D\,,\ r_2=-1/3\,,\,r_1=r_3=1/3\,$ and the metric takes the form
\begin{equation}\label{9}
ds_{_c}=\root 3 \of {(dx_0-dx_1-dx_2-dx_3)(dx_0-dx_1+dx_2+dx_3)(dx_0+dx_1+dx_2-dx_3)}\,.
\end{equation}
On the last (fourth) face $\,A\,C\,D\,,\ r_3=r_1+r_2-1\,$ and we get
\begin{equation}\label{10}
\begin{array}{rl}
ds_{_{ACD}}=&(dx_0+dx_1-dx_2+dx_3)^{(1-r_1)/2}(dx_0-dx_1+dx_2+dx_3)^{(1-r_2)/2}\\
\times &(dx_0-dx_1-dx_2-dx_3)^{(r_1+r_2)/2}\,.\\
\end{array}
\end{equation}
At the central point $\,b\,$ of face $\,A\,C\,D\,,\ r_1=r_2=1/3\,,\,r_3=-1/3\,.$ Therefore,
\begin{equation}\label{11}
ds_{_b}=\root 3 \of {(dx_0-dx_1-dx_2-dx_3)(dx_0-dx_1+dx_2+dx_3)(dx_0+dx_1-dx_2+dx_3)}\,.
\end{equation}		

Let us find out at last what is the form that metric (2) takes on six edges of the tetrahedron $\,A\,B\,C\,D\,,$ starting from edge $\,BD\,.$ According to Fig.\,1, this edge is the intersection of faces $\,A\,B\,D\,$ and $\,B\,C\,D\,.$ Therefore, the relations
\begin{eqnarray*} 
& &1-r_1+r_2-r_3=0\,,\\
& &1+r_1-r_2-r_3=0
\end{eqnarray*}
hold among the parameters $\,r_{\alpha}\,$ at that intersection, whence $\,r_3=1\,, \,r_1=r_2=\tilde{r}\,.$ As a result, we get
\begin{equation}\label{12}
ds_{_{BD}}=\left[\frac{(dx_0-dx_3)-(dx_1+dx_2)}{(dx_0-dx_3)+(dx_1+dx_2)}\right]^{\tilde{r}/2}\sqrt{(dx_0-dx_3)^2-(dx_1+dx_2)^2}\,. 
\end{equation}
In the middle of edge $\,BD\,,\ \,r_3=1\,, \,r_1=r_2=\tilde{r}=0\,,$ and expression (12) reduces to the two-dimensional Minkowski metric $\,ds^2=(dx_0-dx_3)^2-(dx_1+dx_2)^2\,.$

Consider edge $\,AD\,,$ which is the intersection of faces $\,A\,B\,D\,$ and $\,A\,C\,D\,.$ The relations
\begin{eqnarray*} 
& &1-r_1+r_2-r_3=0\,,\\
& &1-r_1-r_2+r_3=0
\end{eqnarray*}
hold among the parameters $\,r_{\alpha}\,$ at that edge, resulting in $\,r_1=1\,, \,r_2=r_3=\tilde{r}\,$	and 
\begin{equation}\label{13}
ds_{_{AD}}=\left[\frac{(dx_0-dx_1)-(dx_2+dx_3)}{(dx_0-dx_1)+(dx_2+dx_3)}\right]^{\tilde{r}/2}\sqrt{(dx_0-dx_1)^2-(dx_2+dx_3)^2}\,. 
\end{equation}		
In the middle of edge $\,AD\,,\ \,r_1=1\,, \,r_2=r_3=\tilde{r}=0\,,$ and expression (13) reduces again to the two-dimensional Minkowski metric $\,ds^2=(dx_0-dx_1)^2-(dx_2+dx_3)^2\,.$

Edge $\,CD\,$ is the intersection of faces $\,A\,C\,D\,$ and $\,B\,C\,D\,.$ At that edge, the relations
\begin{eqnarray*} 
& &1-r_1-r_2+r_3=0\,,\\
& &1+r_1-r_2-r_3=0
\end{eqnarray*}
hold among the parameters $\,r_{\alpha}\,,$ resulting in
$\,r_2=1\,, \,r_1=r_3=\tilde{r}\,.$ Accordingly,
\begin{equation}\label{14}
ds_{_{CD}}=\left[\frac{(dx_0-dx_2)-(dx_1+dx_3)}{(dx_0-dx_2)+(dx_1+dx_3)}\right]^{\tilde{r}/2}\sqrt{(dx_0-dx_2)^2-(dx_1+dx_3)^2}\,. 
\end{equation}
In the middle of edge $\,CD\,,\ \,r_2=1\,, \,r_1=r_3=\tilde{r}=0\,,$ so that\\ $\,ds^2=(dx_0-dx_2)^2-(dx_1+dx_3)^2\,.$

Edge $\,CB\,$ is the intersection of faces $\,A\,B\,C\,$ and $\,B\,C\,D\,.$ At that edge, the relations
\begin{eqnarray*} 
& &1+r_1+r_2+r_3=0\,,\\
& &1+r_1-r_2-r_3=0
\end{eqnarray*}
hold among the parameters $\,r_{\alpha}\,,$ whence $\,r_1=-1\,, \,r_2=-r_3=\tilde{r}\,,$ resulting in
\begin{equation}\label{15}
ds_{_{CB}}=\left[\frac{(dx_0+dx_1)-(dx_2-dx_3)}{(dx_0+dx_1)+(dx_2-dx_3)}\right]^{\tilde{r}/2}\sqrt{(dx_0+dx_1)^2-(dx_2-dx_3)^2}\,. 
\end{equation}
In the middle of edge $\,CB\,,\ \,r_1=-1\,, \,r_2=-r_3=\tilde{r}=0\,$ and\\ $\,ds^2=(dx_0+dx_1)^2-(dx_2-dx_3)^2\,.$

Edge $\,AB\,$ is the intersection of faces $\,A\,B\,C\,$ and $\,A\,B\,D\,.$ At that edge, the relations
\begin{eqnarray*} 
& &1+r_1+r_2+r_3=0\,,\\
& &1-r_1+r_2-r_3=0
\end{eqnarray*}
hold among the parameters $\,r_{\alpha}\,,$ i.e. $\,r_2=-1\,, \,r_1=-r_3=\tilde{r}\,,$ resulting in
\begin{equation}\label{16}
ds_{_{AB}}=\left[\frac{(dx_0+dx_2)-(dx_1-dx_3)}{(dx_0+dx_2)+(dx_1-dx_3)}\right]^{\tilde{r}/2}\sqrt{(dx_0+dx_2)^2-(dx_1-dx_3)^2}\,. 
\end{equation}
In the middle of edge $\,AB\,,\ \,r_2=-1\,, \,r_1=-r_3=\tilde{r}=0\,$ and\\ $\,ds^2=(dx_0+dx_2)^2-(dx_1-dx_3)^2\,.$

The last edge $\,AC\,$ is the intersection of faces $\,A\,B\,C\,$ and $\,A\,C\,D\,.$ At that edge, the relation
\begin{eqnarray*} 
& &1+r_1+r_2+r_3=0\,,\\
& &1-r_1-r_2+r_3=0
\end{eqnarray*}
hold among the parameters $\,r_{\alpha}\,,$ whence $\,r_3=-1\,, \,r_1=-r_2=\tilde{r}\,$ and
\begin{equation}\label{17}
ds_{_{AC}}=\left[\frac{(dx_0+dx_3)-(dx_1-dx_2)}{(dx_0+dx_3)+(dx_1-dx_2)}\right]^{\tilde{r}/2}\sqrt{(dx_0+dx_3)^2-(dx_1-dx_2)^2}\,. 
\end{equation}
In the middle of edge $\,AC\,,\ \,r_3=-1\,, \,r_1=-r_2=\tilde{r}=0\,$ and\\ $\,ds^2=(dx_0+dx_3)^2-(dx_1-dx_2)^2\,.$

The next section will treat the relativistic point mechanics of a particle in the entirely anisotropic space-time (2) using essentially the transformations that constitute the homogeneous 3-parameter noncompact group of isometries of that space-time\,\footnote{As to the respective inhomogeneous 7-parameter group, its Lie algebra can be found in [54]}. By its meaning, the 3-parameter homogeneous isometry group is the relativistic symmetry group for space-time (2) and was found [55] to be Abelian, while its determinant linear transformations are of the form
\begin{equation}\label{18}
x'_i=D\,L_{ik}\,x_k\,,
\end{equation}
where
\begin{equation}\label{19}
D=e^{-(\,r_1\,\alpha _1+r_2\,\alpha _2+r_3\,\alpha _3\,)}\ ,
\end{equation}
$L_{ik}$ designates the unimodular matrices, with
\begin{equation}\label{20}
L_{ik}=\left (
\begin{array}{rrrr}
\cal A&-\cal B&-\cal C&-\cal D\\
-\cal B&\cal A&\cal D&\cal C\\
-\cal C&\cal D&\cal A&\cal B\\
-\cal D&\cal C&\cal B&\cal A\\
\end{array}
\right )\ ,
\end{equation}	
$$
{\cal A}=\cosh \alpha _1\cosh \alpha _2\cosh \alpha _3+
\sinh \alpha _1\sinh \alpha _2\sinh \alpha _3\,,
$$
$$
{\cal B}=\cosh \alpha _1\sinh \alpha _2\sinh \alpha _3+
\sinh \alpha _1\cosh \alpha _2\cosh \alpha _3\,,
$$
$$
{\cal C}=\cosh \alpha _1\sinh \alpha _2\cosh \alpha _3+
\sinh \alpha _1\cosh \alpha _2\sinh \alpha _3\,,
$$
$$
{\cal D}=\cosh \alpha _1\cosh \alpha _2\sinh \alpha _3+
\sinh \alpha _1\sinh \alpha _2\cosh \alpha _3\,\phantom{,}
$$
and $\,\alpha _1\,,\alpha _2\,,\alpha _3\,$ being the group parameters.
The transformations inverse to (18) are of the form
\begin{equation}\label{21}
x_i=D^{-1}\,L^{-1}_{ik}\,x'_k\,,
\end{equation}
where
\begin{equation}\label{22}
L^{-1}_{ik}=\left (
\begin{array}{rrrr}
\tilde{\cal A}&-\tilde{\cal B}&-\tilde{\cal C}&-\tilde{\cal D}\\
-\tilde{\cal B}&\tilde{\cal A}&\tilde{\cal D}&\tilde{\cal C}\\
-\tilde{\cal C}&\tilde{\cal D}&\tilde{\cal A}&\tilde{\cal B}\\
-\tilde{\cal D}&\tilde{\cal C}&\tilde{\cal B}&\tilde{\cal A}\\
\end{array}
\right )\ ,
\end{equation}
\begin{equation}\label{23}
\tilde{\cal A}=\cosh \alpha _1\cosh \alpha _2\cosh \alpha _3-
\sinh \alpha _1\sinh \alpha _2\sinh \alpha _3\,,
\end{equation}
\begin{equation}\label{24}
\tilde{\cal B}=\cosh \alpha _1\sinh \alpha _2\sinh \alpha _3-
\sinh \alpha _1\cosh \alpha _2\cosh \alpha _3\,,
\end{equation}
\begin{equation}\label{25}
\tilde{\cal C}=\sinh \alpha _1\cosh \alpha _2\sinh \alpha _3-
\cosh \alpha _1\sinh \alpha _2\cosh \alpha _3\,,
\end{equation}
\begin{equation}\label{26}
\tilde{\cal D}=\sinh \alpha _1\sinh \alpha _2\cosh \alpha _3-
\cosh \alpha _1\cosh \alpha _2\sinh \alpha _3\,.
\end{equation}
Considering that, similar to Lorentz transformations in  Minkowski space, the transformations (18) link the various inertial reference frames in the Finslerian space (2), it is expedient to replace the group parameters $\ \alpha _1\,,\ \alpha _2\,,\ \alpha _3\ $ with the components $\ v_1=dx_1/dx_0\,,\ v_2=dx_2/dx_0\,,\ v_3=dx_3/dx_0\ $ of the velocity of primed reference frame. Although the respective relations can be found in [55], we shall still present them here:
\begin{eqnarray*}
& &v_1=(\tanh\alpha _1-\tanh\alpha_2\tanh\alpha _3)/(
1-\tanh\alpha_1\tanh\alpha_2\tanh\alpha_3)\,,\\
& &v_2=(\tanh\alpha _2-\tanh\alpha_1\tanh\alpha _3)/(
1-\tanh\alpha_1\tanh\alpha_2\tanh\alpha_3)\,,\\
& &v_3=(\tanh\alpha _3-\tanh\alpha_1\tanh\alpha _2)/(
1-\tanh\alpha_1\tanh\alpha_2\tanh\alpha_3)\,.
\end{eqnarray*}
The inverse relations are
\begin{eqnarray*} 
& &\alpha _1\,=\frac{1}{4}\ln \frac{(1+v_1-v_2+v_3)(1+v_1+v_2-v_3)}
{(1-v_1-v_2-v_3)(1-v_1+v_2+v_3)}\,\,,\\
& &\alpha _2\,=\frac{1}{4}\ln \frac{(1-v_1+v_2+v_3)(1+v_1+v_2-v_3)}
{(1-v_1-v_2-v_3)(1+v_1-v_2+v_3)}\,\,,\\
& &\alpha _3\,=\frac{1}{4}\ln \frac{(1-v_1+v_2+v_3)(1+v_1-v_2+v_3)}
{(1-v_1-v_2-v_3)(1+v_1+v_2-v_3)}\,\,.
\end{eqnarray*}
\vspace*{10mm}

\noindent
{\large{\bf 3\,. Mass shell equation in the entirely anisotropic \\\phantom{my}momentum space}}
\vspace*{5mm}

\noindent 
We shall proceed from the considerations of relativistic invariance and minimality along a straight world line to write an action $\,S\,$ for a free particle in the flat entirely anisotropic  Finslerian space-time (2)\,:
\begin{equation}\label{27}
S=-mc\int\limits_a^b\,ds\,\,,
\end{equation}
where $\,ds\,$ is an interval in the Finslerian space (2). Variation of this action is
\begin{equation}\label{28}
\begin{array}{rl}
\delta S=&-\int\limits_a^b(p_0d\delta x_0-p_1d\delta x_1-p_2d\delta x_2-p_3d\delta x_3)\\
=&(-p_0\delta x_0+p_1\delta x_1+p_2\delta x_2+p_3\delta x_3)\!\!\mid_a^b\\
&+\int\limits_a^b[(dp_0/ds)\delta x_0-(dp_1/ds)\delta x_1-(dp_2/ds)\delta x_2-(dp_3/ds)\delta x_3]ds\,.\\
\end{array}
\end{equation}
If the world line is varied under condition $\,(\delta x_i)\!\!\mid_a=(\delta x_i)\!\!\mid_b=0\,,$ the principle of least action gives $\,p_i=const,\,$ i.e. the rectilinear inertial motion. In turn, variation of the coordinates of point $\,b\,$ under condition $\,p_i=const\,$ gives
\begin{equation}\label{29}
p_0=-\frac{\partial S}{\partial x_0}\,, \quad p_{\alpha}=\frac{\partial S}{\partial x_{\alpha}}\,; \qquad \alpha =1\,, 2\,, 3\,,
\end{equation}
whence it becomes clear the $\,p_i\,$ is a canonical 4-momentum of particle in the Finslerian space (2). Having been expressed via 3-velocity $\,v_{\alpha}=dx_{\alpha}/dx_0\,,$ the  components of the 4-momentum take the form
\begin{eqnarray}\nonumber
p_0=\frac{ds}{dx_0}\left (\frac{dx_0}{ds_{_{B-M}}}\right )^{\!\!4}\{&1-v_1^2-v_2^2-v_3^2-2v_1v_2v_3\,\phantom {r_1r_1[\}\,,}\\\nonumber
+&r_1[(1-v_1^2+v_2^2+v_3^2)v_1+2v_2v_3]\,\phantom {\}\,,}\\\nonumber
+&r_2[(1+v_1^2-v_2^2+v_3^2)v_2+2v_1v_3]\,\phantom {\}\,,}\\\label{30}
+&r_3[(1+v_1^2+v_2^2-v_3^2)v_3+2v_1v_2]\}\,,
\end{eqnarray}
\begin{eqnarray}\nonumber
p_1=\frac{ds}{dx_0}\left (\frac{dx_0}{ds_{_{B-M}}}\right )^{\!\!4}\{&\!\!\!\!\!(1-v_1^2+v_2^2+v_3^2)v_1+2v_2v_3\phantom {+r_1,}\\\nonumber
+&r_1[1-v_1^2-v_2^2-v_3^2-2v_1v_2v_3]\phantom {[rr_2,}\\\nonumber
+&r_2[(1+v_1^2+v_2^2-v_3^2)v_3+2v_1v_2]\phantom {r_2,}\\\label{31}
+&r_3[(1+v_1^2-v_2^2+v_3^2)v_2+2v_1v_3]\}\,,
\end{eqnarray}
\begin{eqnarray}\nonumber
p_2=\frac{ds}{dx_0}\left (\frac{dx_0}{ds_{_{B-M}}}\right )^{\!\!4}\{&\!\!\!\!\!(1+v_1^2-v_2^2+v_3^2)v_2+2v_1v_3\phantom {+r_1,,}\\\nonumber
+&r_1[(1+v_1^2+v_2^2-v_3^2)v_3+2v_1v_2]\phantom {rr_2,}\\\nonumber
+&r_2[1-v_1^2-v_2^2-v_3^2-2v_1v_2v_3]\phantom {[r_1r_2,}\\\label{32}
+&r_3[(1-v_1^2+v_2^2+v_3^2)v_1+2v_2v_3]\}\,,\phantom {\,,}
\end{eqnarray}
\begin{eqnarray}\nonumber
p_3=\frac{ds}{dx_0}\left (\frac{dx_0}{ds_{_{B-M}}}\right )^{\!\!4}\{&\!\!\!\!\!(1+v_1^2+v_2^2-v_3^2)v_3+2v_1v_2\phantom {+rr_1,,}\\\nonumber
+&r_1[(1+v_1^2-v_2^2+v_3^2)v_2+2v_1v_3]\phantom {[r_2r_2,}\\\nonumber
+&r_2[(1-v_1^2+v_2^2+v_3^2)v_1+2v_2v_3]\phantom {[r_1r_2,}\\\label{33}
+&r_3[1-v_1^2-v_2^2-v_3^2-2v_1v_2v_3]\phantom {r_2}\}\,,\phantom {\,,}
\end{eqnarray}
where
\begin{equation}\label{34}
\begin{array}{rl}
\left (dx_0/ds\right )\left (ds_{_{B-M}}/dx_0\right )^{4}=&(1-v_1-v_2-v_3)^{(3-r_1-r_2-r_3)\,/\,4}\\
\times &(1-v_1+v_2+v_3)^{(3-r_1+r_2+r_3)\,/\,4}\\
\times &(1+v_1-v_2+v_3)^{(3+r_1-r_2+r_3)\,/\,4}\\
\times &(1+v_1+v_2-v_3)^{(3+r_1+r_2-r_3)\,/\,4}\,.
\end{array}
\end{equation}	
Here, $\,ds\,$ is metric (2); $\,ds_{_{B-M}}\,$ is the Berwald-Moore metric (3). It should be noted that, starting from formula (30), we put $\,m=c=1\,$ in all the relations.

	According to expressions (30)--(33), four quantities (energy $\,p_0\,$ and 3-momentum $\,p_{\alpha}\,$) are functions of three components, $\,v_{\alpha}\,,$ of particle velocity. The relations (30)--(33), therefore, may be treated to be the equations that determine the parametric form of mass shell, while $\,v_{\alpha}\,$ is taken to be the internal coordinates on that mass shell. We shall demonstrate below that the mass shell equation can be obtained to be an algebraic relation for $\,p_i\,\,.$ As to the associate physical aspect, we see that, just as it should be, the energy $\,p_0\,$ determined by (30) reaches its absolute minimum $\,p_0=1\,$ at $\,v_{\alpha}=0\,,$ i.e. for a particle at rest. However, it is of importance to note that, apart from the rest energy $\,p_0=1\,,$ a particle that resides in the entirely anisotropic space (2) still has a non-zero rest momentum. In virtue of (31)--(33), we get $\,p_1=r_1\,, p_2=r_2\,, p_3=r_3\,\,$ at $\,v_{\alpha}=0\,.$ Moreover, according also to (31)--(33), the 3-momentum direction of the particle differs from its 3-velocity direction, thereby demonstrating that the free particle motion in the entirely anisotropic space is analogous to the motion of a quasiparticle in an entirely anisotropic crystalline medium.
	
Similarly to the case of Minkowski space, the 3-velocity of a particle can be found if its 4-momentum in the entirely anisotropic space is known. To obtain the appropriate respective formulas, we shall start from certain useful intermediate relations that are valid in virtue of (30)--(33), namely,
\begin{equation}\label{35}
\frac{p_0+p_3}{p_1+p_2}=\frac{(1-v_3)(1+r_3)+(v_1+v_2)(r_1+r_2)}{(1-v_3)(r_2+r_3)+(v_1+v_2)(1+r_3)}\,\,,
\end{equation}
\begin{equation}\label{36}
\frac{p_0-p_1}{p_2-p_3}=\frac{(1+v_1)(1-r_1)+(v_2-v_3)(r_2-r_3)}{(1+v_1)(r_2-r_3)+(v_2-v_3)(1-r_1)}\,\,,
\end{equation}
\begin{equation}\label{37}
\frac{p_0+p_1}{p_2+p_3}=\frac{(1-v_1)(1+r_1)+(v_2+v_3)(r_2+r_3)}{(1-v_1)(r_2+r_3)+(v_2+v_3)(1+r_1)}\,\,.
\end{equation}
These relations lead to the following set of three linear equations with respect to $\,v_{\alpha}\,\,:$
\begin{equation}\label{38}
a_{\gamma\alpha}v_{\alpha}=b_{\gamma}\,\,,
\end{equation}	
where	
\begin{equation*}
\begin{array}{rcll}
a_{11}=&&a_{12}&=(p_0+p_3)(1+r_3)-(p_1+p_2)(r_1+r_2)\,,\\
&&&\\
a_{13}=&&b_{1}&=(p_1+p_2)(1+r_3)-(p_0+p_3)(r_2+r_3)\,,\\
&&&\\
a_{21}=&-&b_{2}&=(p_0-p_1)(r_2-r_3)-(p_2-p_3)(1-r_1)\,,\\
&&&\\
a_{22}=&-&a_{23}&=(p_0-p_1)(1-r_1)-(p_2-p_3)(r_2-r_3)\,,\\
&&&\\
a_{31}=&&b_{3}&=(p_2+p_3)(1+r_1)-(p_0+p_1)(r_2+r_3)\,,\\
&&&\\
a_{32}=&&a_{33}&=(p_0+p_1)(1+r_1)-(p_2+p_3)(r_2+r_3)\,.
\end{array}
\end{equation*}
At $\,r_1=r_2=r_3=0\,,$ i.e. in the case of the Berwald-Moore space with metric (3), the set (38) takes the form
\begin{eqnarray}\nonumber
& &(p_0+p_3)v_1+(p_0+p_3)v_2+(p_1+p_2)v_3=(p_1+p_2)\,,\\\nonumber
& &(p_3-p_2)v_1+(p_0-p_1)v_2+(p_1-p_0)v_3=(p_2-p_3)\,,\\\label{39}
& &(p_2+p_3)v_1+(p_0+p_1)v_2+(p_0+p_1)v_3=(p_2+p_3)\,.
\end{eqnarray}
The relations
\begin{equation*}
v_1=\frac{p_1({p_0}^2-{p_1}^2+{p_2}^2+{p_3}^2)-2p_0p_2p_3}{p_0({p_0}^2-{p_1}^2-{p_2}^2-{p_3}^2)+2p_1p_2p_3}\,,
\end{equation*}
\begin{equation*}
v_2=\frac{p_2({p_0}^2+{p_1}^2-{p_2}^2+{p_3}^2)-2p_0p_1p_3}{p_0({p_0}^2-{p_1}^2-{p_2}^2-{p_3}^2)+2p_1p_2p_3}\,,
\end{equation*}
\begin{equation*}
v_3=\frac{p_3({p_0}^2+{p_1}^2+{p_2}^2-{p_3}^2)-2p_0p_1p_2}{p_0({p_0}^2-{p_1}^2-{p_2}^2-{p_3}^2)+2p_1p_2p_3}
\end{equation*}
solve the set (39).

As noted above, four functions, (30)--(33), of three variables $\,v_{\alpha}\,$ determine the parametric form of mass shell. Let now obtain the algebraic form of the mass shell equation, i.e. $\,H^4(p_0,\,p_1,\,p_2,\,p_3)=1\,.$ The explicit form of the function $\,H^4(p_0,\,p_1,\,p_2,\,p_3)\,$ can be found as follows. First, four relations, which are valid in virtue of (30)--(33), are to be written\,:
\begin{equation}\label{40}
\frac{p_0+p_1+p_2+p_3}{1+r_1+r_2+r_3}=\frac{ds}{dx_0}\left (\frac{dx_0}{ds_{_{B-M}}}\right )^{\!\!4}(1-v_1+v_2+v_3)\left [(1+v_1)^2-(v_2-v_3)^2\right ]\,\,,
\end{equation}
\begin{equation}\label{41}
\frac{p_0+p_1-p_2-p_3}{1+r_1-r_2-r_3}=\frac{ds}{dx_0}\left (\frac{dx_0}{ds_{_{B-M}}}\right )^{\!\!4}(1-v_1-v_2-v_3)\left [(1+v_1)^2-(v_2-v_3)^2\right ]\,\,,
\end{equation}
\begin{equation}\label{42}
\frac{p_0-p_1+p_2-p_3}{1-r_1+r_2-r_3}=\frac{ds}{dx_0}\left (\frac{dx_0}{ds_{_{B-M}}}\right )^{\!\!4}(1+v_1+v_2-v_3)\left [(1-v_1)^2-(v_2+v_3)^2\right ]\,\,,
\end{equation}
\begin{equation}\label{43}
\frac{p_0-p_1-p_2+p_3}{1-r_1-r_2+r_3}=\frac{ds}{dx_0}\left (\frac{dx_0}{ds_{_{B-M}}}\right )^{\!\!4}(1+v_1-v_2+v_3)\left [(1-v_1)^2-(v_2+v_3)^2\right ]\,\,.
\end{equation}
After that, examine the structure of the expressions in the right-hand parts of (40)--(43). Considering the structure demonstrated by formula (34) for the common factor $\,\left (dx_0/ds\right )\left (ds_{_{B-M}}/dx_0\right )^{4}\,,$ it can readily be noted that the right-hand parts of (40)-(43) are the products of different powers of four characteristic ``brackets" $\,(1-v_1-v_2-v_3),\ \\
(1-v_1+v_2+v_3),\ (1+v_1-v_2+v_3)\,$ and $\,(1+v_1+v_2-v_3)\,.$ This observation suggests that the function $\,H^4(p_0,\,p_1,\,p_2,\,p_3)\,$ should be sought for as
\begin{eqnarray}\nonumber
H^4(p_0,\,p_1,\,p_2,\,p_3)&=&\left (\frac{p_0+p_1+p_2+p_3}{1+r_1+r_2+r_3}\right )^{\!a}\left (\frac{p_0+p_1-p_2-p_3}{1+r_1-r_2-r_3}\right )^{\!b}\\\label{44}
&\times &\left (\frac{p_0-p_1+p_2-p_3}{1-r_1+r_2-r_3}\right )^{\!c}\left (\frac{p_0-p_1-p_2+p_3}{1-r_1-r_2+r_3}\right )^{\!d}\,\,.
\end{eqnarray}
The first of the conditions to be imposed on the constants $\,a\,,b\,,c\,$ and $\,d\,$ ensues from the physical meaning of the function $\,H(p_0,\,p_1,\,p_2,\,p_3)\,$ and consists in that the function must have a physical dimension that coincides with the dimension of momentum $\,p_i\,.$ Therefore, the function (44) should be a homogeneous function of its own arguments of the fourth degree of homogeneity. This requirement means that
\begin{equation}\label{45}
a+b+c+d=4\,.
\end{equation}
The rest conditions to be imposed on the constants $\,a\,,b\,,c\,$ and $\,d\,$ can be obtained in terms of the requirement that all the power exponents, which arise for four characteristic ``brackets" after substituting the expressions (40)--(43) in (44), should equal zero. It is just in this case that we obtain the mass shell equation in the form $\,H^4(p_0,\,p_1,\,p_2,\,p_3)=1\,.$ Considering, however, that we have put $\,m=c=1\,,$ the equation $\,H^4(p_0,\,p_1,\,p_2,\,p_3)=1\,$ corresponds in the ordinary units to $\,H^4(p_0,\,p_1,\,p_2,\,p_3)=(mc)^4\,$. 

So, if the proposed program is fulfilled, then we get the following four equations for the constants $\,a\,,b\,,c\,$ and $\,d\,$ to supplement the equation (45)\,:
\begin{eqnarray}\label{46}
& &b+c+d-(3-r_1-r_2-r_3)(a+b+c+d)/4=0\,,\\\label{47}
& &a+c+d-(3-r_1+r_2+r_3)(a+b+c+d)/4=0\,,\\\label{48}
& &a+b+d-(3+r_1-r_2+r_3)(a+b+c+d)/4=0\,,\\\label{49}
& &a+b+c-(3+r_1+r_2-r_3)(a+b+c+d)/4=0\,.
\end{eqnarray}		
In virtue of (45), the set of five equations (45)--(49) can be rewritten as
\begin{eqnarray}\label{50}
& &a+b+c+d=4\,,\\\label{51}
& &b+c+d-(3-r_1-r_2-r_3)=0\,,\\\label{52}
& &a+c+d-(3-r_1+r_2+r_3)=0\,,\\\label{53}
& &a+b+d-(3+r_1-r_2+r_3)=0\,,\\\label{54}
& &a+b+c-(3+r_1+r_2-r_3)=0\,.
\end{eqnarray}	
Obviously, we get (50) by summing up equations (51)--(54). Therefore, (50) is not an independent equation, so four independent equations (51)--(54), or the respective set
\begin{eqnarray}\label{55}
& &b+c+d=(3-r_1-r_2-r_3)\,,\\\label{56}
& &a+c+d=(3-r_1+r_2+r_3)\,,\\\label{57}
& &a+b+d=(3+r_1-r_2+r_3)\,,\\\label{58}
& &a+b+c=(3+r_1+r_2-r_3)
\end{eqnarray}
remain to determine four constants $\,a\,,b\,,c\,$ and $\,d\,.$\\ The constants
$$
\begin{array}{ll}
a=1+r_1+r_2+r_3\,,&b=1+r_1-r_2-r_3\,,\\
c=1-r_1+r_2-r_3\,,&d=1-r_1-r_2+r_3\\
\end{array}
$$								
solve the set (55)--(58). This result means that the equation of mass shell in the entirely anisotropic momentum space is
\begin{eqnarray}\nonumber
&&\left (\frac{p_0+p_1+p_2+p_3}{1+r_1+r_2+r_3}\right )^{\!(1+r_1+r_2+r_3)}
\left (\frac{p_0+p_1-p_2-p_3}{1+r_1-r_2-r_3}\right )^{\!(1+r_1-r_2-r_3)}\\\label{59}
&\times&\left (\frac{p_0-p_1+p_2-p_3}{1-r_1+r_2-r_3}\right )^{\!(1-r_1+r_2-r_3)}
\left (\frac{p_0-p_1-p_2+p_3}{1-r_1-r_2+r_3}\right )^{\!(1-r_1-r_2+r_3)}=1\,.
\end{eqnarray}

Finally, we shall consider the relativistic symmetry group of the entirely anisotropic momentum space and show that the transformations of the 4-momenta that form the group leave the mass shell equation (59) invariant. From the general considerations it becomes clear that the transformations of relativistic symmetry of the entirely anisotropic momentum space are induced by the respective transformations (18) of the entirely anisotropic event space (2). The explicit form of the linear transformations of 4-momenta that represent the relativistic symmetry group will be constructed proceeding from the definition of the canonical 4-momentum (29).

Thus, the relations
\begin{equation}\label{60}
p'_0=-\,\frac{\partial S}{\partial x_i}\,\frac{\partial x_i}{\partial x'_0}=D^{-1}(\,L^{-1}_{00}\,p_0-L^{-1}_{0\beta}\,p_{\beta}\,)\,,
\end{equation}
\begin{equation}\label{61}
p'_{\alpha}=\frac{\partial S}{\partial x_i}\,\frac{\partial x_i}{\partial x'_{\alpha}}=D^{-1}(\,-L^{-1}_{\alpha 0}\,p_0+L^{-1}_{\alpha\beta}\,p_{\beta}\,)
\end{equation}
are valid in virtue of (29) and (21). Considering the definition (22) of matrix $\,L^{-1}_{ik}\,,$ we can unite the relations (60) and (61) into a single formula
\begin{equation}\label{62}
p'_i=D^{-1}\,{\cal L}_{ik}\,p_k\,,
\end{equation}
where
\begin{equation}\label{63}
D^{-1}=e^{(\,r_1\,\alpha _1+r_2\,\alpha _2+r_3\,\alpha _3\,)}\ ,
\end{equation}
\begin{equation}\label{64}
{\cal L}_{ik}=\left (
\begin{array}{rrrr}
\tilde{\cal A}&\tilde{\cal B}&\tilde{\cal C}&\tilde{\cal D}\\
\tilde{\cal B}&\tilde{\cal A}&\tilde{\cal D}&\tilde{\cal C}\\
\tilde{\cal C}&\tilde{\cal D}&\tilde{\cal A}&\tilde{\cal B}\\
\tilde{\cal D}&\tilde{\cal C}&\tilde{\cal B}&\tilde{\cal A}\\
\end{array}
\right )
\end{equation}	
in virtue of (19) and (22). Here, $\ \alpha _1\,,\ \alpha _2\,,\ \alpha _3\ $ are the group parameters; the matrix elements $\,\tilde{\cal A}\,,\,\tilde{\cal B}\,,\,\tilde{\cal C}\,$ and $\,\tilde{\cal D}\,$ of the matrix $\,{\cal L}_{ik}\,$ are determined by formulas (23)--(26). Thus, we have constructed the explicit form (62) of the linear transformations of 4-momenta. The transformations give rise to the 3-parameter Abelian group of relativistic symmetry of the entirely anisotropic momentum space.

To verify that the form of the mass shell equation (59) is actually invariable, i.e. remains invariant, under transformations (62), it is expedient to find out first in what way four independent 1-forms entering equation (59) are transformed. The straightforward calculations by (62)--(64) and by (23)--(26) give
\begin{equation}\label{65}
(p'_0+p'_1+p'_2+p'_3)=e^{[\,(r_1-1)\,\alpha _1+(r_2-1)\,\alpha _2+(r_3-1)\,\alpha _3\,]}(p_0+p_1+p_2+p_3)\,,
\end{equation}
\begin{equation}\label{66}
(p'_0+p'_1-p'_2-p'_3)=e^{[\,(r_1-1)\,\alpha _1+(r_2+1)\,\alpha _2+(r_3+1)\,\alpha _3\,]}(p_0+p_1-p_2-p_3)\,,
\end{equation}
\begin{equation}\label{67}
(p'_0-p'_1+p'_2-p'_3)=e^{[\,(r_1+1)\,\alpha _1+(r_2-1)\,\alpha _2+(r_3+1)\,\alpha _3\,]}(p_0-p_1+p_2-p_3)\,,
\end{equation}
\begin{equation}\label{68}
(p'_0-p'_1-p'_2+p'_3)=e^{[\,(r_1+1)\,\alpha _1+(r_2+1)\,\alpha _2+(r_3-1)\,\alpha _3\,]}(p_0-p_1-p_2+p_3)\,.
\end{equation}
Thus, as should be expected, the transformations of the relativistic symmetry of the entirely anisotropic momentum space get much simplified in terms of 1-forms and reduce only to the scale transformations of those independent forms. Using (65)--(68), it can readily be verified that
\begin{eqnarray}\nonumber
&&(p'_0+p'_1+p'_2+p'_3)^{\!(1+r_1+r_2+r_3)}
(p'_0+p'_1-p'_2-p'_3)^{\!(1+r_1-r_2-r_3)}\\\nonumber
&\times&(p'_0-p'_1+p'_2-p'_3)^{\!(1-r_1+r_2-r_3)}
(p'_0-p'_1-p'_2+p'_3)^{\!(1-r_1-r_2+r_3)}\\\nonumber
&=&(p_0+p_1+p_2+p_3)^{\!(1+r_1+r_2+r_3)}
(p_0+p_1-p_2-p_3)^{\!(1+r_1-r_2-r_3)}\\\nonumber
&\times&(p_0-p_1+p_2-p_3)^{\!(1-r_1+r_2-r_3)}
(p_0-p_1-p_2+p_3)^{\!(1-r_1-r_2+r_3)}\,\,.
\end{eqnarray}
It is this equality that proves that equation (59) remains invariant under transformations (62).

Obviously, the equation (59) describes the 3-parameter family of mass shells in the entirely anisotropic momentum space. Therefore, (59) can be used as a generatrix, thereby relating each of the possible mass shells to a respective entirely anisotropic metric.
\vspace*{12mm}

\noindent
{\large{\bf 4\,. Conclusion}}
\vspace*{6mm}

\noindent
Having studied all the possible metric states of the flat entirely anisotropic space-time and on supplementing them with the state that corresponds to the partially violated isotropy, we obtained the total ensemble of the relativistically-invariant metric states of the Finslerian event space and, respectively, the entire spectrum of geometric phase transitions from the absolute time to 4D space-time. Since each of such transitions is associated with a certain channel of spontaneous violation of the initial gauge symmetry and with the structure of the appearing condensate, the above result gets determinant in the course of the Finslerian Standard Model Extension (FSME).

The FSME is attractive because it permits the relativistic symmetry, represented now by the group of generalized Lorentz boosts, to be preserved. As to Lorentz symmetry, it proves to be violated. However, the violation is described by a much smaller (compared with SME) number of parameters. They are just the parameters that determine the structure of the condensates formed under spontaneous violation of the initial gauge symmetry.

	As regards its structure,  condensate may prove to be either axially symmetric or entirely anisotropic to one or another extent, with the classification of the entirely anisotropic condensates coinciding with the classification of the flat entirely anisotropic  Finslerian metrics realized in the present work. From the constructive viewpoint, however, it is of importance that any of the above mentioned condensates can be obtained as a constant, i.e  $(t,x,y,z)$--independent solution of the respective nonlinear generalized field equation.
	
 To avoid any misunderstanding, we note finally that, for example, the standard Dirac equation does not have any condensate-like solution, while its nonlinear generalization, which satisfies the requirement of the generalized Lorentz, i.e DISIM${_{b}}$(2) invariance, admits such a solution (see in this connection [19] and [56]).
\newpage
\noindent
{\large{\bf Acknowledgment}}
\vspace*{3mm}

\noindent
The author is grateful to Prof. Igor Volobuev for valuable discussion. This work is supported in part by the Russian Foundation for Basic Research
under grant RFBR 07-01-91681.

\end{document}